\documentclass[a4paper, reqno]{amsart}
\usepackage{a4}
\usepackage{amsmath,graphicx, amssymb,color,ulem}
\usepackage[sort&compress,numbers]{natbib}

\usepackage{setspace}

\newtheorem{theorem}{Theorem}

\newtheorem{proposition}[theorem]{Proposition}

\newcommand{\ud}{\mathrm{d}}

\begin{document}

\title[Net reproduction rate of populations with distributed states at
birth]{On the net reproduction rate of continuous structured populations with distributed
states at birth}

\author[A.\,S.  Ackleh]{Azmy S. Ackleh}
\address{Azmy S. Ackleh, Department of Mathematics, University of Louisiana at Lafayette, Lafayette, LA 70504, USA}
\email{ackleh@louisiana.edu}

\author[J.\,Z. Farkas]{J\'{o}zsef Z. Farkas}
\address{J\'{o}zsef Z. Farkas, Division of Computing Science and Mathematics, University of Stirling, Stirling, FK9 4LA, United Kingdom }
\email{jozsef.farkas@stir.ac.uk}

\date{\today}

\begin{abstract}
\begin{sloppypar}
We consider a nonlinear structured population model with a distributed recruitment term.
The question of the existence of non-trivial steady states can be treated (at least) in three
different ways. One approach is to study spectral pro\-per\-ti\-es of a parametrised family of unbounded operators.
The alternative approach, which we develop here, is based on the reformulation of the partial differential e\-qu\-ation as an integral equation. In this context we introduce a density dependent net reproduction rate and discuss its relationship to a biologically meaningful quantity. Finally, we discuss a third approach, which is based on a finite rank approximation of the recruitment operator.

\end{sloppypar}
\end{abstract}
\maketitle

\section{Introduction}

In mathematical epidemiology the basic reproductive number, often
denoted by $R_0$, is the expected number of secondary infections that a single
infected individual will cause in a completely susceptible population when there
is no control or intervention, see e.g.~\cite{NI}. In the mathematical context
these ideal assumptions mean basically that the underlying model is a linear
one (or indeed it is a linear approximation of a nonlinear one). Hence it is expected that the infection will persist if $R_0>1$. It is also
expected that larger values of $R_0$ will result in a major epidemic and
significant intervention efforts will be needed to control the spread of the
disease. In a mathematical model $R_0$ is in fact often introduced as a
threshold parameter that allows one to determine the stability of the disease free steady state,
the spectral radius of a positive bounded linear operator, see \cite{DHM}.

In population ecology/dynamics, the value $R_0$ is often related to the
measure of reproductive success of an individual in an ideal environment, see
e.g.~\cite{CUS}. It can be  used as a bifurcation parameter to study the
existence and local asymptotic stability (in some cases global asymptotic stability) of the extinction (or in some cases the positive) steady state.
In the context of the following linear age-structured population model
\begin{equation}\label{linage}
p_t(a,t)+p_a(a,t)=-\mu(a)p(a,t),\quad p(0,t)=\int_0^m \beta(a)p(a,t)\,\ud a,\quad p(a,0)=p_0(a),
\end{equation}
one introduces the quantity
\begin{equation*}
R=\int_0^m\beta(a)\exp\left\{-\int_0^a\mu(r)\,\ud r\right\}\,\ud a,
\end{equation*}
which is the expected number of newborns of an individual to be produced in her
lifetime.  It is then shown that $p(a,t)\sim e^{rt}p_*(a)$, as $t\to\infty$,
where $r$ and $R-1$ have the same sign (see \cite{HT} for general results of this type).
This asymptotic property of solutions is
called asynchronous exponential growth, and it can be effectively
characterized in the framework of semigroup theory, see e.g. \cite{AGG,CH,W}.
We also note that, equivalently, $R$ can be defined as the spectral radius of the
Volterra integral operator defined by the right hand side of the following renewal equation
\begin{equation}\label{renewal}
B(t)=\int_0^m\beta(a)\exp\left\{-\int_0^a\mu(r)\,\ud r\right\}B(t-a)\,\ud a ,
\end{equation}
for the density of newborns $B(t)=p(0,t)$. The advantage of formulating the linear
age-structured problem \eqref{linage} as a renewal equation is that integral
operators are usually nicely behaved (for example bounded) in contrast with unbounded differential operators.

As the above simple linear example illustrates, net reproduction numbers are
expected to play a key role in the analysis of both discrete and continuous
population models. Recently, this has been a topic of interest. In
the context of structured epidemiological models we refer the reader to the
recent papers by Baca\"{e}r at al. \cite{B,BD,BO},  while in the context of
discrete structured population models we refer to \cite{CUS,CUS2,CA}.
It is clear that if one incorporates nonlinearities into the simple age-structured model
\eqref{linage}, it cannot be expected that a simple constant will determine the
asymptotic behaviour of solutions, for example whether the population will
persist or die out, in general. However, it is still true that questions of existence and
local asymptotic stability properties of steady states can naturally be related
to appropriately defined net reproduction functions (or functionals), see for example \cite{JZF,FH}.

In the simple age-structured model \eqref{linage} above it is quite natural that recruitment of individuals into the population
takes place at age $0$, as they represent the newborns. In models however, when the population is structured with respect to size or any other phy\-si\-o\-lo\-gi\-cal variable, it is not clear why recruitment should (only) take place at the minimal size. In fact there are many concrete applications in which it is clear that a single state at birth model cannot be a good approximation of the problem. For example this is the case in several cell populations in which reproduction is by fission and the size of the daughter cell is not fixed, and it is determined for example by a pro\-ba\-bi\-li\-ty distribution function, see e.g. \cite{Heijmans1} and Section III in \cite{MD}. To take this effect into account one introduces a recruitment operator, usually in the form of a bounded integral operator, in the partial differential equation. This introduces some complications in the mathematical analysis but at the same time one gets rid of the non-local and often non-linear boundary condition which describes the influx at the minimal size (or state).

Recently, in \cite{FGH} we considered the following non-linear Gurtin-MacCamy-type model (see \cite{GM}), with a
distributed recruitment term.
\begin{align}
\frac{\partial}{\partial t}p(s,t)+\frac{\partial}{\partial
s}\left(\gamma(s,P(t))p(s,t)\right)&=-\mu(s,P(t))p(s,t)+\int_0^1\beta(s,y,
P(t))p(y,t)\,\ud y,\label{equation} \\
\gamma(0,P(t))p(0,t)&=0,\label{boundary} \\
p(s,0)&=p_0(s),\ \ P(t)=\int_0^1 p(s,t)\,\ud s\label{initial}.
\end{align}
Here we set the maximal size to be $1$ for mathematical convenience.
In this model it is assumed that individuals may have different sizes at birth
and therefore  $\beta(s,y,\,\cdot\,)$ denotes the rate at which individuals of
size $y$ give rise to individuals of size $s$. Hence the non-local integral term
in equation \eqref{equation} represents reproduction of the population without external
driving of the population through immigration. Population models with distributed recruitment processes
have been applied for example to model cell populations and cell aggregation problems, see e.g. \cite{MD}. Integral operators representing distributed recruitment/loss processes also appear in physical problems, for example in the modelling of coagulation-fragmentation processes (see e.g., \cite{Ackleh,AF,MLB2,MLB}).

We make the following regularity assumptions on the model ingredients, which are needed in what follows.
\begin{align}
& \gamma\in C^2([0,1]\times[0,\infty)),\,\, \mu\in C^1([0,1]\times [0,\infty)),\,\, \beta\in C^1([0,1]\times [0,1]\times [0,\infty)), \nonumber \\
& 0<\gamma_0\le\gamma\le\Gamma,\quad 0\le\mu\le M,\quad 0\le\beta\le B.\label{regconditions}
\end{align}
Some of the above conditions are biologically relevant, such as the assumption of non-negative and bounded vital rates,
the others, such as differentiability, are necessary to discuss linear stability of steady states, see later in Section 5.
Existence of solutions of the model above (and in fact of a much more general model) was treated in
\cite{CS2}. It is relatively straightforward to verify that the regularity conditions \eqref{regconditions} we impose on the model ingredients guarantee that hypotheses (A1)-(A4) in \cite{CS2} (which are needed for the proof of the existence result to hold true) are satisfied. We also note that in \cite{CS2} the authors in fact treated the much more delicate case, when
the size-space may be unbounded.

In \cite{FGH} we established some sufficient conditions for the existence of a positive steady state of this
model. We summarize here the key steps of the analysis we employed in \cite{FGH}, mainly to understand the differences and similarities with the new developments in the next Section. For a fixed $P\in [0,\infty)$ one defines a linear operator $\mathcal{A}_P$ by
\begin{align}
\mathcal{A}_P\,u= & -\frac{\partial}{\partial
s}\left(\gamma(\cdot,P)u\right)-\mu(\cdot,P)u+\int_0^1\beta(\cdot,y,P)u(y)\,\ud
y, \nonumber \\
\text{Dom}(\mathcal{A}_P)= & \left\{u\in W^{1,1}(0,1)\,|\,u(0)=0\right\}. \label{spectproblem}
\end{align}
Then one needs to establish conditions which guarantee that there exists a $P_*$ such that
$\mathcal{A}_{P_*}$ has eigenvalue $0$ with a corresponding unique
po\-si\-ti\-ve eigenvector. (See also \cite{CS} where this approach was employed in case of a cyclin-structured cell population model.) To this end, we established that $\mathcal{A}_{P}$ is the generator of a positive and irreducible (under some mild condition on $\beta$, see later in Section 3) semigroup. In
\cite{FGH} we also established that the semigroup generated by $\mathcal{A}_P$
is eventually compact, which implies that the Spectral Mapping Theorem holds
true, and the spectrum of $\mathcal{A}_P$ may contain only isolated eigenvalues
of finite algebraic multiplicity (see e.g. \cite{NAG}). It then follows that the
spectral bound is a dominant (real) eigenvalue $\lambda_P=s\left(\mathcal{A}_P\right)$ of geometric
multiplicity one with a corresponding po\-si\-ti\-ve eigenvector, see e.g. \cite[Chapter
9]{CH}. Finally, we established conditions which imply that there exist a
$P^+\in (0,\infty)$ such that the spectral bound $s(\mathcal{A}_{P^+})$ is
negative and therefore the dominant eigenvalue
$\lambda_{P^+}=s(\mathcal{A}_{P^+})$ is also negative; and a $P^-\in (0,\infty)$
such that this dominant eigenvalue $\lambda_{P^-}= s(\mathcal{A}_{P^-})$ is
positive. Then it follows from standard perturbation results on eigenvalues (see
e.g.~\cite{K}) that there exists a zero eigenvalue, and there is a corresponding positive eigenvector, which then
is normalised to obtain a positive steady state.

The reformulation of the steady state problem as an eigenvalue problem for a
family of unbounded linear operators allowed us to obtain sufficient
conditions for the existence of positive steady states. Unfortunately, the
biological motivation is somewhat lost in the approach above. In the present
paper we treat the question of existence of positive steady states
of model \eqref{equation}-\eqref{initial} using a different method. Our
motivation is  to establish the existence of positive steady states via an
appropriate net reproduction function as in the case of the basic age- (or size-) structured Gurtin-MacCamy model
with one state at birth. Nevertheless, since local (and in some cases global)
asymptotic stability of equilibria can naturally be related to an appropriate net reproduction
function,  see e.g. \cite{JZF, FGH, FH}.

\section{Existence of steady states and the net reproduction function}

In this section we discuss the existence of steady states of model \eqref{equation}-\eqref{initial} by  first (re)formulating the steady state problem as an integral equation.
To this end, we introduce the function
\begin{equation}\label{inteqF}
B^*(s):=B(p_*,s)=\int_0^1\beta(s,y,P_*)p_*(y)\,\ud y.
\end{equation}
With this notation, an implicit solution of the steady state equation can be obtained as
\begin{equation}\label{steadystate}
p_*(y)=\frac{1}{\gamma(y,P_*)}\int_0^y\exp\left\{-\int_r^y\frac{\mu(x,P_*)}{\gamma(x,P_*)}\,\ud x\right\}B^*(r)\,\ud r.
\end{equation}
This yields the integral equation
\begin{equation}\label{inteqB}
B^*(s)=\int_0^1\frac{\beta(s,y,P_*)}{\gamma(y,P_*)}\int_0^y\exp\left\{-\int_r^y\frac{\mu(x,P_*)}{\gamma(x,P_*)}\,\ud x\right\}B^*(r)\,\ud r\,\ud y,\quad s\in [0,1].
\end{equation}
Using the notation
\begin{equation*}
K(s,r,P_*)=\int_r^1\exp\left\{-\int_r^y\frac{\mu(x,P_*)}{\gamma(x,P_*)}\,\ud x\right\}\frac{\beta(s,y,P_*)}{\gamma(y,P_*)}\,\ud y,
\end{equation*}
and a change of variables in the integration, we can recast equation \eqref{inteqB} in the more economic form:
\begin{equation}\label{inteqB2}
B^*(s)=\int_0^1K(s,r,P_*)B^*(r)\,\ud r,\quad s\in [0,1].
\end{equation}
The existence of a non-trivial (and non-negative) solution of equation \eqref{inteqB2} is necessary for the existence of a non-trivial steady state $p_*$.
On the other hand, once a non-trivial solution of \eqref{inteqB2} is found we can substitute this directly into \eqref{steadystate} to get a positive steady state.
Also note that if the integral equation \eqref{inteqB2} has a non-negative solution, then any positive scalar multiple of that is a solution, hence  we can normalize the function $B^*$ such that the steady state will also satisfy the following necessary condition:
\begin{equation*}
P_*=\int_0^1 p_*(y)\,\ud y.
\end{equation*}

We note that problem $\eqref{inteqB2}$ can be treated as an eigenvalue problem, but now for a bounded operator. In particular for a fixed ``environment'' $P$ we consider the integral operator defined as
\begin{equation}\label{intoperator}
\mathcal{L}_P x=\int_0^1 K(\cdot,r,P)x(r)\,\ud r, \quad \text{for} \quad x\in\mathcal{X}=L^1(0,1).
\end{equation}
More precisely, we consider a family of integral operators parametrised by $P$.
The existence of a positive (not necessarily strictly positive) steady state requires that for some $P_*>0$ the operator $\mathcal{L}_{P_*}$ has eigenvalue one with a corresponding positive (not necessarily strictly positive) eigenvector $x$.

The relationship between problems \eqref{spectproblem} and \eqref{intoperator} is established using the main result of \cite{HT},
which we recall here for the reader's convenience.

\begin{theorem} (\cite[Theorem 3.5]{HT})
Let $\mathcal{B}$ be a resolvent-positive operator in $\mathcal{X}$, $s(\mathcal{B})<0$, and $\mathcal{A} =\mathcal{B}+\mathcal{C}$ a positive perturbation of $\mathcal{B}$. If $\mathcal{A}$ is resolvent-positive then $s(\mathcal{A})$ has the same sign as $r\left(-\mathcal{C}\,\mathcal{B}^{-1}\right)-1$.
\end{theorem}
In our setting for every $P\in [0,\infty)$ we have $\mathcal{A}_P=\mathcal{B}_P+\mathcal{C}_P$, where $\mathcal{A}_P$ is defined in \eqref{spectproblem} and
\begin{align}
\mathcal{B}_P\,u= & -\frac{\partial}{\partial
s}\left(\gamma(\cdot,P)u\right)-\mu(\cdot,P)u, \quad \text{Dom}(\mathcal{B}_P)=\left\{u\in W^{1,1}(0,1)\,|\,u(0)=0\right\},  \label{operatorB} \\
\mathcal{C}_P\,u= &\int_0^1\beta(\cdot,y,P)u(y)\,\ud y, \quad \text{Dom}(\mathcal{C}_P)=\mathcal{X}. \label{operatorC}
\end{align}
It is clear, that for every $P\in[0,\infty)$, $\mathcal{C}_P$ is a positive operator and $\mathcal{B}_P$ is resolvent-positive,
i.e. its resolvent set contains a positive half-line and its resolvent $(\lambda\mathcal{I}-\mathcal{B}_P)^{-1}$ is
positive for $\lambda$ large enough.  This is because it is shown that it generates a positive quasi-contractive semigroup (see \cite{FGH}).
We also note that if $\displaystyle\inf\{\mu(s,P)\,|\, s\in [0,m]\}\ge\nu>0$ holds then it is shown that $\mathcal{B}_P$ generates a positive contraction semigroup, in particular $\omega_0(\mathcal{B}_P)=s(\mathcal{B}_P)\le -\nu$.
We note that the assumption of a strictly positive mortality function is very natural from the biological point of view for most populations.
Next we show that for every $P\in  [0,\infty)$ the integral operator $\mathcal{L}_P$ defined in \eqref{intoperator} is in fact $-\mathcal{C}_P\mathcal{B}_P^{-1}$. Since we have
\begin{align*}
& \frac{\ud}{\ud y}\left( \int_0^y\exp\left\{-\int_r^y\frac{\mu(x,P_*)}{\gamma(x,P_*)}\,\ud x\right\}u(r)\,\ud r \right) \\
& \quad\quad=u(y)-\frac{\mu(y,P)}{\gamma(y,P)}\int_0^y\exp\left\{-\int_r^y\frac{\mu(x,P_*)}{\gamma(x,P_*)}\,\ud x\right\}u(r)\,\ud r,
\end{align*}
 it is easily shown, that if we define an operator $\mathcal{B}^{-1}_P$ as
\begin{equation}
\mathcal{B}^{-1}_P\, u=\frac{-1}{\gamma(\cdot,P)}\int_0^\cdot\exp\left\{-\int_r^\cdot\frac{\mu(x,P)}{\gamma(x,P)}\,\ud x\right\}u(r)\,\ud r
\end{equation}
with $\text{Dom}\left(\mathcal{B}^{-1}_P\right)=\mathcal{X}$,  then we have
\begin{equation}
\mathcal{B}_P\,\mathcal{B}_P^{-1}\, u=u,
\end{equation}
for every $P\in [0,\infty)$, hence $\mathcal{B}^{-1}_P$ is indeed the (right) inverse of $\mathcal{B}_P$.
If the regularity conditions \eqref{regconditions} hold true then it is shown that the operator
$\mathcal{B}_P^{-1}$ maps the state space $\mathcal{X}$ into the domain of $\mathcal{B}_P$ for every $P$. Thus, we have
\begin{equation}
\mathcal{L}_P\, u=-\mathcal{C_P}\mathcal{B}^{-1}_P\, u,
\end{equation}
for all $u\in\mathcal{X}$ and $P\in [0,\infty)$.

We also note that problem \eqref{intoperator} can be treated as a ``fixed-ray" problem in the positive cone of
$\mathcal{X}$. In particular it is clear that the operator $\mathcal{L}_P$ (restricted to the positive cone of $\mathcal{X}$)
maps positive rays into positive rays, and any fixed ray yields an eigenvalue $\lambda_P$,
where $\mathcal{L}_P$ restricted to this fixed ray is a multiplication operator with $\lambda_P$.
The idea of discussing existence of non-trivial steady states in the framework of a combination of fixed point results and spectral theory is very fruitful, see e.g. \cite{CF,Pal,FHin2}.

\begin{theorem}\label{nonemptyspectrum}
For every $P\in [0,\infty)$ the spectrum of $\mathcal{L}_P$ is not empty and contains only eigenvalues of finite multiplicity.
\end{theorem}
\noindent {\bf Proof.}
Since the kernel $K$ is continuous for every $P\in [0,\infty)$, the linear integral operator $\mathcal{L}_P$ is bounded
for every $P\in [0,\infty)$ with
\begin{equation*}
||\mathcal{L}_P||_\infty=\max_{x\in [0,1]}\int_0^1 K(x,y,P)\,\ud y,
\end{equation*}
hence it is continuous and compact.

We define a map $\Phi\,:\,B^+_1\to B^+_1$ as $\Phi=\mathcal{P}\circ\mathcal{L}_P$, where
$B^+_1=\{x\in\mathcal{X}_+\,|\,||x||=1\}$ the unit sphere intersected with the positive cone of $\mathcal{X}$ and $\mathcal{P}$ is a (continuous) projection onto $B^+_1$ along positive rays in $\mathcal{X}$.
Note that  the set $B^+_1$ is convex since $\mathcal{X}=L^1$ is an AL-space. The Banach lattice $\mathcal{X}$
is called an abstract L-space (AL space) if $||f+g||=||f||+||g||$ for $f,g\ge 0$, see e.g. \cite{AGG}.
We apply Schauder's fixed point theorem to the compact map $\Phi$. A fixed point of this map $\Phi$ in turn implies the existence of a
fixed-ray $R=\left\{\alpha\,x\,\vert\,x\in B^+_1,\,\alpha\ge 0\right\}$ of the integral operator $\mathcal{L}_P$. On this ray $R$ the operator $\mathcal{L}_P$
acts as a multiplication operator with a constant say $\lambda_P$, i.e. $\mathcal{L}_P x=\lambda_P x$. This $\lambda_P$ is therefore an eigenvalue of $\mathcal{L}_P$ with a corresponding eigenvector which is any (non-zero) element of the fixed-ray $R$.
 \hfill $\Box$

In what follows we discuss spectral properties of the operator $\mathcal{L}_P$. For basic concepts and results not introduced here we refer to \cite{AGG,KR,Sch}. An important consequence of Theorem \ref{nonemptyspectrum} and Theorem 2.9 in \cite{KR} is that for every $P\in [0,\infty)$ the spectral radius $r(\mathcal{L}_P)$ is a positive eigenvalue with a corresponding positive eigenvector. For more recent related developments we refer the interested reader to \cite{HT2}. Note that the proof of Theorem \ref{nonemptyspectrum} shows that there is a positive eigenvalue of $\mathcal{L}_P$  which has a positive eigenvector, but it does not imply that the spectral radius has a corresponding positive eigenvector.

Next recall for example from \cite{Marek} that a (non-trivial) continuous positive endomorphism $\mathcal{O}$ on the Banach lattice $\mathcal{X}$ is called (ideal) irreducible if it does not admit closed invariant ideals other than the trivial ones. We note that it is well-known (see e.g. \cite[Sect.V.6]{Sch}) that a compact and positive irreducible operator on an AL-space has non-empty point spectrum. However, in Theorem \ref{nonemptyspectrum} we did not assume that $\mathcal{L}_P$ is irreducible.

Since for every $P\in [0,\infty)$ $\mathcal{L}_P$ is compact and positive the spectral radius $r(\mathcal{L}_P)$ is an eigenvalue, i.e. it is a pole of the resolvent of $\mathcal{L}_P$. Hence it follows, see e.g. \cite[Sect.V.5]{Sch}, that if $\mathcal{L}_P$ is irreducible then the spectral radius $r(\mathcal{L}_P)$ is the only eigenvalue with a positive (and strictly positive) eigenvector. On the other hand if $\mathcal{L}_P$ is not irreducible then there may be other eigenvalues in the spectrum which admit positive (not necessarily strictly positive) eigenvectors.

Therefore, if the operator $\mathcal{L}_P$ is irreducible for every $P$, then it is natural to define the net reproduction rate of the standing population to be the spectral radius of $\mathcal{L}_P$, i.e.  $R(P):=r\left(\mathcal{L}_P\right)$.  If for example $\beta$ is strictly positive, then this is clearly the case. See later in Section 3 the characterisation of irreducibility of $\mathcal{L}_P$. However, if $\beta$ vanishes on some set of positive measure then $\mathcal{L}_P$ is not necessarily irreducible.
For example, if there is a maximal offspring size m (less than the maximal size), then the ideal
of equivalence classes of $L^1$ functions vanishing on $(m,1)$ will be invariant under $\mathcal{L}_P$.
Note that in this case we cannot shrink the state space to the interval $(0,m)$, since
individuals of size greater than $m$ may still reproduce.
Most importantly, if there is another eigenvalue $\lambda_P$ of the operator $\mathcal{L}_P$
with a corresponding positive eigenvector $x_P$ (or fixed-ray) which belongs to one of those ideals,
i.e. it vanishes on $(m,1)$, then this eigenvector still yields a positive (and even strictly positive if it does not vanish on some interval $(0,\varepsilon)$ for $\varepsilon>0$) steady state via formula \eqref{steadystate}. Hence this (or for that matter any other) eigenvalue could be considered as a net reproduction rate.

The idea of proving existence of positive steady states via the net reproduction function works now
exactly as in the case of simple Gurtin-McCamy models.
Let us write:
\begin{equation}
\sigma\left(\mathcal{L}_P\right)=\left\{\lambda^1_P, \lambda^2_P, \cdots, \lambda^i_P,\cdots|\, i\in N\subseteq\mathbb{N} \right\}.
\end{equation}
If $\lambda^i_P$ is any eigenvalue such that $\lambda_0^i>1$ and $\displaystyle\lim_{P\to\infty}\lambda_P^i=0$ then it follows from the
continuous dependence of the eigenvalue on the parameter $P$ (see e.g. \cite{K}) that there exists a $P_*>0$ such that
$\lambda^i_{P_*}=1$. Also, if $\lambda_{P_*}^i$ admits a positive (not necessarily strictly positive if $i\ne 1$) normalised eigenvector $x_*$ then this is a fixed ray for the operator $\mathcal{L}_P$. We can then multiply this eigenvector $x_*$ with an appropriate positive constant (and write $B_*$ for this new vector) such that
\begin{equation*}
P_*=\int_0^1\frac{1}{\gamma(y,P_*)}\int_0^y\exp\left\{-\int_r^y\frac{\mu(x,P_*)}{\gamma(x,P_*)}\,\ud x\right\} B_*(r)\,\ud r\ud y
\end{equation*}
holds.
Then, formula \eqref{steadystate} yields a non-trivial steady state.

\section{Results for special types of kernels}

As we noted in the previous section, from the mathematical point of view a natural candidate for the net reproduction rate, at least in the case when the integral operator $\mathcal{L}_P$ is irreducible, would be the spectral radius $r(\mathcal{L}_P)$. From the biological point of view however a natural candidate is the following function
\begin{equation}\label{netreprate}
R(P)=\int_0^1\int_0^1K(s,r,P)\,\ud r\,\ud s.
\end{equation}
This is because the kernel $K$ gives the average density of individuals of size $s$ produced by individuals who were born at size $r$,
if the standing population is $P$ (and would remain constant $P$).
In this section we show that the definition of the net reproduction function discussed in the previous section
(defined via eigenvalues of an integral operator) coincides with the biologically meaningful net reproduction function
$R(P)$ defined in \eqref{netreprate} at least in the case of special classes of kernels of the integral operator $\mathcal{L}_P$  defined in \eqref{intoperator}.
\begin{theorem}\label{equality2}
If the kernel $K$ satisfies
\begin{equation}\label{netreprate3}
C_P=\int_0^1K(s,r,P)\,\ud r,\quad s\in [0,1],\quad P\in [0,\infty),
\end{equation} then for every $P\in [0,\infty)$,
\begin{equation}
r\left(\mathcal{L}_P\right)=\int_0^1\int_0^1K(s,r,P)\,\ud r\,\ud s=C_P=R(P).
\end{equation}
\end{theorem}
\noindent {\bf Proof.}
First we show that if \eqref{netreprate3} holds then $R(P)\le r\left(\mathcal{L}_P\right)$.
To this end, we are going to utilize a minimax principle from \cite{Marek} .
Recall from \cite{Marek} that a subset $H'\subseteq K'$ is called $K$-total if and only if from the relations $\langle x,x'\rangle\ge 0$ for all $x'\in H'$ it follows that $x\in K$.
Note that, in our setting $K'$ is $K$-total, since $K$ is closed. For any $x\in K$ define
\begin{equation*}
r_x(T)=\sup_{w}\left\{w\in\mathbb{R}\,|\,(Tx-wx)\in K\right\}.
\end{equation*}
Lemma 3.1 in \cite{Marek}  states that if $H'\subseteq K'$ is $K$-total then
\begin{equation*}
r_{x}(T)=\sup_{v}\left\{v\in\mathbb{R}\,|\,\langle Tx,x'\rangle\ge v\langle x,x'\rangle,\, x'\in H'\right\}.
\end{equation*}
Also recall from \cite{Marek} (Lemma 3.3) that if $T$ is a bounded linear positive endomorphism of $\mathcal{X}$ and $K$ is closed
then for any $0\ne x\in K$ we have
\begin{equation}\label{spectralineq}
r_{x}(T)\le r(T).
\end{equation}
We choose $x=1$, for which we have for every $x'\in K'$
\begin{equation*}
\langle \mathcal{L}_P\,1,x'\rangle\ge C_P\int_0^1x'(s)\,\ud s.
\end{equation*}
Hence for every $P\in [0,\infty)$ we have
\begin{equation*}
R(P)\le r_1(\mathcal{L}_P)\le r(\mathcal{L}_P).
\end{equation*}

To show the inequality $r(\mathcal{L}_P)\le R(P)$ we utilise again a result from \cite{Marek}. Recall that for a positive bounded linear endomorphism $T$ we have
\begin{equation}
s^{x'}(T)=\inf_{\tau}\left\{\tau\in\mathbb{R}\,|\, \tau\langle x,x'\rangle\ge \langle Tx,x'\rangle,\, x\in K\right\}.
\end{equation}
Again, let us choose $x'=1$ and it is easy to see that $C_P\langle x,1\rangle\ge \langle \mathcal{L}_P,1\rangle$ for every $x\in K$,
hence we have
\begin{equation}
r(\mathcal{L}_P)\le s^1(\mathcal{L}_P)\le R(P).
\end{equation}
\hfill $\Box$

\begin{remark}
From the biological point of view condition \eqref{netreprate3} implies that individuals produce the same amount of offspring of different sizes during their lifetime. A simple example of a fertility function for  which condition \eqref{netreprate3} holds is  $\beta(s,y,P)= \beta(y,P)$, i.e, individuals of size $y$ produce individuals of any size $s$ at the same rate (thus $\beta$ is independent of $s$). However, individuals of different sizes $y$ at different population levels $P$
may have different fertility rates (thus the dependency of $\beta$ on $y$ and $P$).
\end{remark}

\begin{remark}\label{sizeindep}
We note that the dual statement of Theorem \ref{equality2} can be proven similarly. That is, if there exists a function $C^*$ such that
\begin{equation}\label{netreprate2}
C^*_P=\int_0^1 K(s,r,P)\,\ud s,\quad \forall r\in [0,1],
\end{equation}
then $R(P)=r(\mathcal{L}_P)$. However, condition \eqref{netreprate2} cannot be satisfied for meaningful vital rates.
\end{remark}

Next we return to the case when the integral operator is irreducible, since in this case as we have shown earlier, the spectral radius $r(\mathcal{L}_P)$ is the only eigenvalue with a positive eigenvector.
Recall from \cite{Sch} that the integral operator $\mathcal{L}_P$ is irreducible if and only if for every $S\subset [0,1]$ (where $S$ has positive Lebesgue measure) we have
\begin{equation}\label{irredcond}
\displaystyle\int\limits_{[0,1]\setminus S}\int\limits_{S} K(x,y,P)\,\ud y\,\ud x>0.
\end{equation}
Note that, for example $\beta>0$ implies \eqref{irredcond}. We denote by $\mathcal{L}^*_P$ the adjoint of $\mathcal{L}_P$. For an irreducible integral operator we obtain the following result.
\begin{proposition}\label{minmax}
If the kernel $K$ satisfies \eqref{irredcond} for every $P\in [0,\infty)$ then
\begin{align}
& \min_y\int_0^1 K(x,y,P)\,\ud x\le r(\mathcal{L}_P)\le \max_y\int_0^1K(x,y,P)\,\ud x,\label{minmax1} \\
& \min_x\int_0^1 K(x,y,P)\,\ud y\le r(\mathcal{L}^*_P)\le \max_x\int_0^1K(x,y,P)\,\ud y.\label{minmax2}
\end{align}
\end{proposition}
\noindent {\bf Proof.}
Since $\mathcal{L}_P$ is irreducible, let $f\in\mathcal{X}_+$ denote the strictly positive eigenvector corresponding to the spectral radius
$r(\mathcal{L}_P)$. We have
\begin{equation}\label{mineq1}
\int_0^1K(x,y,P)f(y)\,\ud y=r(\mathcal{L}_P)f(x),\quad x\in [0,1].
\end{equation}
Integration of \eqref{mineq1} and Fubini's Theorem yields:
\begin{equation}\label{mineq2}
\int_0^1f(y)\int_0^1 K(x,y,P)\,\ud x\,\ud y=r(\mathcal{L}_P)\int_0^1f(x)\,\ud x.
\end{equation}
From \eqref{mineq2} we immediately obtain \eqref{minmax1} noting that the kernel $K$ is positive.

Next we note that $r(\mathcal{L}_P)=r(\mathcal{L}^*_P)$, and that if $K$ satisfies \eqref{irredcond} then $\mathcal{L}^*_P$ is
also irreducible. We proceed similarly as above to obtain the inequalities \eqref{minmax2} for $r(\mathcal{L}^*_P)$.
\hfill $\Box$

\begin{remark}
Note that the inequalities \eqref{minmax1}-\eqref{minmax2} together with the assumption of Theorem \ref{equality2} would trivially imply that $r(\mathcal{L}_P)=R(P)$. However, Proposition \ref{minmax}
only holds for an irreducible operator and the condition of Theorem \ref{equality2} does not
imply that $\mathcal{L}_P$ is irreducible. In particular for the simplest case of fertility function  $\beta(s,y,P)=\beta(y,P)$ for which \eqref{netreprate3} holds, the irreducibility condition \eqref{irredcond} requires:
\begin{equation}\label{irredcond2}
\displaystyle\int\limits_S\int_r^1\exp\left\{-\int_r^y\frac{\mu(x,P)}{\gamma(x,P)}\,\ud x\right\}\frac{\beta(y,P)}{\gamma(y,P)}\,\ud y\,\ud r>0,
\end{equation}
for any subset $S\subset [0,1]$ of positive Lebesgue measure. It is shown that condition \eqref{irredcond2} holds if $\exists\,\varepsilon>0$ such that $\beta(y,P)>0$ for $y\in [1-\varepsilon,1]$. In the context of a size-structured model this means that individuals of the largest sizes still produce offspring.
Let us point out that the irreducibility condition \eqref{irredcond} (for the integral operator $\mathcal{L}_P$) is much stronger than that of the semigroup generated by $\mathcal{A}_P$ for a general recruitment function $\beta$. It is shown that the semigroup generated by $\mathcal{A}_P$ is irreducible if $\exists\,\varepsilon>0$ such that $\beta(s,y,P)>0$ for $s\in [0,\varepsilon]$
and $y\in [1-\varepsilon,1]$ for $P\in [0,\infty)$, see \cite{FGH}. This condition however does not imply \eqref{irredcond}, as the following simple example shows. Let $\beta(s,y,P)$ be a function which is positive in some neighbourhood of the point
$(s=0,y=1)$ for every $P\in [0,\infty)$, and vanishes for all $y\le \frac{3}{4},\,s\in [0,1],\,P\in [0,\infty)$. Moreover, let $S=\left[0,\frac{1}{4}\right]\cup\left[\frac{3}{4},1\right]$. Then it is clear that \eqref{irredcond}  cannot hold.
\end{remark}

\begin{remark}
We note that one cannot expect $r(\mathcal{L}_P)=\int_0^1\int_0^1 K(x,y,P)\,\ud y\,\ud x$ to hold in general,
since it does not even hold for all (irreducible) $2\times 2$ matrices.
\end{remark}

\begin{remark} In the special case of a separable fertility function $\beta$, e.g. if $\beta(s,y,P)=\beta_1(s)\beta_2(y,P)$ for $s,y\in [0,1],\,P\in [0,\infty)$ for some functions $\beta_1,\beta_2$,
one can define a net reproduction function, which can be related to the existence of positive steady states of model \eqref{equation}-\eqref{initial}, see \cite{FGH}. In fact in this case the operator $\mathcal{L}_P$ is $\beta_1$-positive
(and $\mathcal{X}_+$ is reproducing; see \cite{KR} for the definitions) hence Theorem 2.11 in \cite{KR} guarantees that
$\mathcal{L}_P$ has only one positive eigenvector.
\end{remark}

\section{Finite rank approximation of the recruitment term}

In this section we briefly outline another approach to treat the steady state problem, which relies on the approximation of the recruitment operator with finite rank operators. As we noted before, the general model \eqref{equation}-\eqref{initial} cannot be solved explicitly even for a time independent solution. Instead, we consider the approximate problem with a fertility function $\beta^n$ defined as
\begin{equation*}
\beta^n(s,y,P)=\displaystyle\sum_{r=1}^n\beta_r(s)\bar{\beta}
_r(y)\tilde{\beta} _r(P),
\end{equation*}
with $\beta_r,\bar{\beta}_r,\tilde{\beta}\in C^0$.
We note that for this type of separable fertility function the recruitment integral operator is of finite
rank (at most $n$). Hence for this fertility function $\beta^n$ we find the solution (explicitly) of the steady state equation as
\begin{equation}\label{steadystate2}
p^n_*(s)=\sum_{r=1}^n Q_*^{r,n}\tilde{\beta}_r(P^n_*) F_r(s,P^n_*),
\end{equation}
where
\begin{align*}
F_r(s,P)&=\int_0^s\exp\left\{-\int_y^s\frac{\mu(x,P)+\gamma_s(x,P)}{\gamma(x,P)
} \, \ud x\right\}\frac{\beta_r(y)}{\gamma(y,P)}\,\ud y \\
&= \int_0^s\exp\left\{-\int_y^s\frac{\mu(x,P)}{\gamma(x,P)}\, \ud
x\right\}\frac{\beta_r(y)}{\gamma(s,P)}\,\ud y,
\end{align*}
and
\begin{equation*}
Q_*^{r,n}=\int_0^1\bar{\beta}_r(s)p^n_*(s)\,\ud s,\quad P^n_*=\int_0^1p^n_*(s)\,\ud s.
\end{equation*}
Multiplying equation \eqref{steadystate2} by $\bar{\beta}_j(s)$ and integrating from $0$
to $1$ we obtain an $n$-dimensional system
\begin{equation*}
\mathbf{Q}_*^n=\mathbf{O}^n_{P^n_*}\mathbf{Q}_*^n,
\end{equation*}
where $\mathbf{Q}_*^n=(Q_*^{1,n},\cdots,Q_*^{n,n})^T$ is an $n$-dimensional vector for every $n$ and $\mathbf{O}^n_{P}$ is an $n\times n$  matrix valued function with positive elements $o_{ij}^n(P)$, where
\begin{equation*}
o_{ij}^n(P)=\tilde{\beta}_i(P)\int_0^1\bar{\beta}_i(s)F_j(s,P)\,\ud s,\quad i,j=1,\dots,n.
\end{equation*}
It can be shown that a positive stationary solution to the approximate $\beta^n$-problem exists if and
only if there exists a $P_*>0$ such that the matrix $\mathbf{O}^n_{P_*}$ has
eigenvalue one with a corresponding strictly positive eigenvector.
Note that similarly to the general case we need to assure that, now a finite-dimensional operator,
has eigenvalue $1$ with a positive eigenvector.
Hence we may call $\mathbf{O}^n_P$ (purely motivated from the mathematical point of view) the net reproduction matrix (corresponding to the approximate $\beta^n$ problem). Clearly, $\mathbf{O}^n(P)=
\mathbf{O}^n_P$ is a continuous function of the variable $P$ if $\mu,\gamma$
and the $\tilde{\beta}$'s are continuous functions of $P$.

The spectral radius of the matrix $\mathbf{O}^n_P$ is a monotone decreasing function of $P$ if the elements
$o^n_{ij}(P)$ are monotone decreasing functions of $P$. Hence for any fixed $n$ one can establish conditions on the model ingredients that guarantee the existence of
a value $P$ such that the corresponding matrix $\mathbf{O}^n_P$ has
spectral radius $1$. It follows then from Perron-Frobenius theory that this is an eigenvalue with a corresponding strictly positive eigenvector.
This eigenvector then yields via formula \eqref{steadystate2} a strictly positive steady state of the approximate problem.

Hence from the mathematical point of view it seems reasonable for the approximate problem to define the net reproduction rate of the standing population as the spectral radius of the matrix $\mathbf{O}^n_P$.
Next one may ask naturally the question whether a limiting linear operator-valued function exists,
that is, one would like to show that
\begin{equation*}
\lim_{n\to\infty}\mathbf{O}^n_P=\mathcal{O}_P,
\end{equation*}
with convergence for example with respect to the topology induced by the operator norm.
This question is outside the scope of the present paper and is left for future work. We note however, that this
limiting operator (if it exists) admits all the nice properties as does the integral operator $\mathcal{L}_P$.
Firstly, because of the uniform convergence in $P$, the limiting operator is a continuous function of $P$.
$\mathcal{O}_P$ is also compact since it is a limit of
operators of finite rank. Positivity of $\mathcal{O}_P$ follows immediately from the fact that the
positive cone of $l^1$ is closed. That is, Krein-Rutman theory (the infinite dimensional analogue of Perron-Frobenius theory) may be applied to study spectral properties of $\mathcal{O}_P$. The relationship however, between this parametrised family of limiting operators $\mathcal{O}_P$ (acting naturally
on $l^1$), and the family of integral operators $\mathcal{L}_P$ (acting on $L^1$) defined in \eqref{intoperator} is far from clear. We note that, to discuss the relationship between these two objects one would naturally need to embed the two state spaces into a Banach space of measures.

The finite rank approximation of the fertility function may be useful for concrete applications.
In principle, for any finite rank approximation of $\beta$ (and given model ingredients) one can compute the spectral radius
of the matrix valued function $\mathbf{O}^n_P$, and this way obtain an approximate net reproduction function of the standing
population, which can then be used to predict qualitative behaviour of the model.

\section{Concluding remarks}

In this paper we investigated a nonlinear structured population model with distributed states at birth.
We defined a biologically meaningful density dependent net reproduction rate, see formula \eqref{netreprate}.
We also related this net reproduction rate to a mathematically well grounded quantity, namely the spectral radius of a linear integral operator.
This is motivated by the question of the existence of positive steady states, see also \cite{DHM, DGM}.
We have established some conditions which guarantee that these two quantities equal. We also presented another approach,
which is based on the approximation of the fertility rate in the original problem. This approximation leads then to a finite dimensional problem, namely to an eigenvalue problem for a matrix, which is more tractable.

It is apparent from the recent literature, see e.g. \cite{B,BD,BD2,BO,Inaba,NI} that there is a recurrent interest in mathematical approaches defining net reproductive numbers for epidemic models. This is partly due to the number of epidemic outbreaks of a variety of diseases in the 21st century. We point out that most of the authors have been focusing on non-autonomous but linear models. We also note that distributed recruitment terms, such as the one we considered here, naturally appear in epidemic models, see e.g. \cite{CF}, where the structuring variable represents pathogen load.

From the ecological or population dynamic point of view intrinsic growth rates and net reproductive numbers play a key role in the qualitative analysis of deterministic models. They are often used as bifurcation parameters
when proving the existence of positive steady states, see e.g. \cite{CUS,CUS2,CA}.
It is also often the case, that the value of the net reproduction function at $0$, i.e. $R(0)$
(or $R_0$) determines the local asymptotic stability of the extinction steady state. This is also the case
at least for Gurtin-MacCamy-type nonlinear structured population models, see e.g. \cite{FH}.
There is also a possibility that if the net reproduction rate at the zero population density, i.e. $R_0$, is greater than one, then the positive steady state of the model is (globally) asymptotically stable.
The idea behind this is that since nonlinearities often represent competition effects,
it is natural to assume that the density dependent net reproduction rate is a monotone decreasing function of the
standing population size. Unfortunately, even if this is the case, there is still the possibility that the positive steady state looses its stability via Hopf-bifurcation, and periodic (stable) cycles emerge.
It is however the case, see e.g. \cite{JZF,FH,FHin2}, that density dependent net reproduction
functions (or functionals) play a key role in the local stability analysis of positive steady states
of physiologically structured population dynamic models.

Similarly to the the single state at birth model it is also possible to relate the value $R(0)$ to the local asymptotic stability of the extinction
equilibrium. We recall from \cite{FGH} that the linearised system around the trivial steady state reads:
\begin{align}
& u_t(s,t)=-\gamma(s,0)\,u_s(s,t)-\left(\gamma_s(s,0)+ \mu(s,0)\right)\,u(s,t)+\int_0^1\beta(s,y,0)u(y,t)\,\ud y,\nonumber\\
& \gamma(0,0)u(0,t)=0,\quad U(t)=\int_0^1 u(s,t)\,\ud s.\label{lin}
\end{align}
The governing linear semigroup is eventually compact, see \cite{FGH}, hence it is enough to consider the eigenvalue equation when
determining stability conditions for the extinction steady state. This reads:
\begin{equation}\label{eigenproblem}
-\gamma(s,0)\,v'(s)-\left(\gamma_s(s,0)+ \mu(s,0)\right)\,v(s)+\int_0^1\beta(s,y,0)v(y)\,\ud y=\lambda v(s),\,\,v(0)=0.
\end{equation}
The solution of equation \eqref{eigenproblem} is found (if it exists) as
\begin{equation}\label{eigensol}
v(s)=\int_0^s\frac{f(y)}{\gamma(y,0)}\exp\left\{-\int_y^s\frac{\lambda+\mu(r,0)+\gamma'(r,0)}{\gamma(r,0)}\,\ud r\right\}\,\ud y,
\end{equation}
where we introduced the notation
\begin{equation*}
f(x)=\int_0^1\beta(x,y,0)v(y)\,\ud y.
\end{equation*}
Hence $\lambda\in\mathbb{C}$ is an eigenvalue if and only if equation \eqref{eigensol} has a nontrivial solution $v$.
Mulitplying equation \eqref{eigensol} by $\beta$ and integrating from $0$ to $1$ yields a similar integral equation to
\eqref{inteqB}, the characteristic equation of the linearisation at the extinction steady state. Thich reads:
\begin{equation}\label{eigeninteq}
f(x)=\int_0^1\frac{\beta(x,s,0)}{\gamma(s,0)}\int_0^yf(y)\exp\left\{-\int_y^s\frac{\lambda+\mu(r,0)}{\gamma(r,0)}\,\ud r\right\}\,\ud y\,\ud s,\quad x\in [0,1].
\end{equation}
In particular, for $\lambda=0$, equation \eqref{eigeninteq} is identical to problem \eqref{inteqB} whith $P_*=0$.
We may now define a family of integral operators parametrised by $\lambda$ (at least for real $\lambda$) $\mathcal{N}_\lambda$ where
$\mathcal{N}_\lambda f$ is defined via the right hand side of equation \eqref{eigeninteq}.
For every $\lambda$ this operator is positive, moreover its spectral radius $r(\mathcal{N}_\lambda)$ (which is an eigenvalue)
is a monotone decreasing function of $\lambda$. We therefore conclude, that at least when the kernel $K$ (or the net reproduction function $R$) is a monotone decreasing function of $P$, then the existence of a positive steady state implies that equation \eqref{eigeninteq} has a solution with $\lambda>0$, hence the trivial steady state is unstable. Hence we showed, that at least in special cases the definition of $R$ agrees with the intuitive understanding that the net
reproductive number $R_0$ detemines local asymptotic stability of the extinction steady state and stability may be lost with the emergence of positive steady state. Here we only briefly considered the relationship between the net reproduction number and the stability of the trivial steady state. In \cite{FGH} we established some linear stability
results for strictly positive steady states. In \cite{BIMV} linear stability analysis of a similar Gurtin-McCamy type model was presented. That approach is based on the discretization of the linearised operator. On the other hand, we shall note that the principle of linearised stability for quasilinear equations such as the one we discussed here has only been established in \cite{GH} for the case of a separable growth rate $\gamma$. Hence future efforts may focus on the development of
an accurate numerical method to solve the non-linear equations and establish convergence of the numerical solutions.
The numerical method will then be used to verify for more cases of the kernel whether such a net reproductive value being less
than one results in population extinction and if it is larger than one then the population persists.

\section*{Acknowledgments}
A.~S. Ackleh thanks the Edinburgh Mathematical Society for partial support while visiting the University of Stirling.
J.~Z. Farkas was supported by a University of Stirling Research and Enterprise Support Grant and a Royal Society International Grant.
Much of this work was done when J.~Z. Farkas visited the University of Lousiana at Lafayette.
Financial support and the hospitality of the Department of Mathematics is greatly appreciated.
We thank the referees for their valuable comments.


\begin{thebibliography}{99}

\bibitem{Ackleh}
A.S. Ackleh, Estimation of parameters in a structured algal coagulation-fragmentation model, {\em Nonlin. Anal. TMA}, {\bf 28} (1997), 837-854.

\bibitem{AF}
A.S. Ackleh and B.G. Fitzpatrick,
Modeling aggregation and growth processes in an algal population model: analysis and computation,
{\em J. Math. Biol.}, {\bf 35} (1997), 480-502.

\bibitem{AGG}
W. Arendt, A. Grabosch, G. Greiner, U. Groh, H. P. Lotz, U. Moustakas, R. Nagel, F. Neubrander and U. Schlotterbeck,
\newblock {\em One-Parameter Semigroups of Positive Operators},
\newblock Springer-Verlag, Berlin, (1986).

\bibitem{B}
N. Baca\"{e}r,
\newblock Periodic matrix population models: growth rate, basic reproduction number, and entropy,
\newblock  {\em Bull. Math. Biol.}, {\bf 71} (2009), 1781-1792.

\bibitem{BD}
N. Baca\"{e}r and E. H. A. Dads,
\newblock On the biological interpretation of a definition for the parameter $R_0$ in periodic population models,
\newblock  {\em J. Math. Biol.}, {\bf 65} (2012),  601-621.

\bibitem{BD2}
N. Baca\"{e}r and E. H. A. Dads,
\newblock Genealogy with seasonality, the basic reproduction number, and the influenza pandemic,
\newblock  {\em J. Math. Biol.}, {\bf 62} (2011),  741-762.

\bibitem{BO}
N. Baca\"{e}r and R. Ouifki,
\newblock Growth rate and basic reproduction number for population models with a simple periodic factor,
\newblock  {\em Math. Biosci.}, {\bf 210} (2007),  647-658.

\bibitem{CS}
R. Borges, \`A. Calsina, S. Cuadrado,
\newblock Equilibria of a cyclin structured cell population model,
\newblock {\em Discrete Contin. Dyn. Syst., Ser. B},  {\bf 11} (2009),  613-627.

\bibitem{BIMV}
D. Breda, M. Iannelli, S. Maset, and R. Vermiglio,
\newblock Stability analysis of the Gurtin-MacCamy model,
\newblock {\em SIAM J. Numer. Anal.}, {\bf 46} (2008), 980-995.

\bibitem{CF}
\`A. Calsina and J.~Z. Farkas,
\newblock Steady states in a structured epidemic model with Wentzell boundary condition,
\newblock {\em J. Evol. Equ.}, {\bf 12}  (2012), 495-512.

\bibitem{Pal}
\`A. Calsina and  J. M. Palmada,
Steady states of a selection-mutation model for an age structured population,
\newblock {\em J. Math. Anal. Appl.}, {\bf 400}  (2013), 386-395.

\bibitem{CS2}
\`A. Calsina and J. Salda\~na,
\newblock Basic theory for a class of models of hierarchically structured
population dynamics with distributed states in the recruitment,
\newblock {\em Math. Models Methods Appl. Sci.}\  {\bf 16} (2006), 1695-1722.

\bibitem{CH}
Ph. Cl\'{e}ment, H.\,J.\,A.\,M Heijmans, S. Angenent, C.\,J. van Duijn, and B. de Pagter,
\newblock {\em One-Parameter Semigroups},
\newblock North--Holland, Amsterdam 1987.

\bibitem{CUS}
J. M. Cushing,
\newblock {\em An introduction to structured population dynamics},
\newblock SIAM, Philadelphia, 1998.

\bibitem{CUS2}
J. M. Cushing,
\newblock On the relationship between $r$ and $R_0$ and its role in the bifurcation of stable equilibria of Darwinian matrix models,
\newblock { \em J. Biol. Dyn.},  {\bf 5} (2011), 277-297.

\bibitem{CA}
J. M. Cushing and A. S. Ackleh,
\newblock A net reproductive number for periodic matrix models,
\newblock { \em J. Biol. Dyn.}, {\bf 6} (2012), 166-188.

\bibitem{DHM}
O. Diekmann, J. A. P. Heesterbeek and J. A. J. Metz,
\newblock On the definition and the computation of the basic reproduction ratio $R_0$ in models for infectious diseases in heterogeneous populations,
\newblock { \em J. Math. Biol.}, {\bf 28} (1990), 365-382.

\bibitem{DGM}
O. Diekmann, M. Gyllenberg and J. A. J. Metz,
\newblock Steady-state analysis of structured population models,
\newblock { \em Theoret. Population Biol.}, {\bf 63} (2003), 309-338.

\bibitem{NAG}
K.-J. Engel, R. Nagel.
One-Parameter Semigroups for Linear Evolution Equations.
Springer, New York, 2000.

\bibitem{JZF}
J. Z. Farkas,
\newblock Size-structured populations: immigration, (bi)stability and the net growth rate,
\newblock  {\em J. Appl. Math. Comput.} {\bf 35} (2011),  617-633.

\bibitem{FGH}
J. Z. Farkas, D. Green and P. Hinow,
\newblock Semigroup analysis of structured parasite populations,
\newblock  {\em Math. Model. Nat. Phenom.}, {\bf 5} (2010), 94-114.

\bibitem{FH}
J. Z. Farkas, T. Hagen,
\newblock Stability and regularity results for a size-structured population model,
\newblock {\em J. Math. Anal. Appl.}, {\bf 328} (2007), 119--136.

\bibitem{FHin2}
J. Z. Farkas and P. Hinow,
\newblock Steady states in hierarchical structured populations with distributed states at birth,
\newblock   {\em Discrete Contin. Dyn. Syst. Ser. B},  {\bf 17} (2012), 2671-2689.

\bibitem{GH}
A. Grabosch and H.~J.~A.~M. Heijmans. \newblock Cauchy problems with
state-dependent time evolution, \newblock {\em Japan J. Appl. Math.}\ {\bf 7}
(1990), 433--457.

\bibitem{GM}
M. E. Gurtin and R. C. MacCamy,
\newblock Non-linear age-dependent population dynamics,
\newblock {\em Arch. Rat. Mech. Anal.}, {\bf 54} (1974), 281-300.

\bibitem{Heijmans1}
H. J. A. M. Heijmans,
\newblock On the stable size distribution of populations reproducing by fission into two unequal parts,
\newblock {\em Math. Biosci.}, {\bf 72}  (1984), 19-50.

\bibitem{Heijmans2}
H. J. A. M. Heijmans,
\newblock An eigenvalue problem related to cell growth,
\newblock {\em J. Math. Anal. Appl.}, {\bf 111}  (1985), 253-280.

\bibitem{Inaba}
 H. Inaba,
\newblock On a new perspective of the basic reproduction number in heterogeneous environments,
\newblock  {\em J. Math. Biol.}, {\bf 65} (2012), 309-348.

\bibitem{K}
T. Kato,
\newblock {\em Perturbation Theory for Linear Operators},
\newblock Springer, New York, (1966).

\bibitem{KR}
M. A. Krasnoselʹskii,
\newblock {\em Positive solutions of operator equations},
\newblock P. Noordhoff Ltd., Groningen, (1964).

\bibitem{MLB2}
D. J. McLaughlin, W. Lamb, and A. C. McBride,
\newblock An existence and uniqueness result for a coagulation and multiple-fragmentation equation,
\newblock {\em SIAM J. Math. Anal.}, {\bf 28} (1997), 1173–1190.

\bibitem{MLB}
D. J. McLaughlin, W. Lamb, and A. C. McBride,
\newblock A semigroup approach to fragmentation models,
\newblock {\em SIAM J. Math. Anal.}, {\bf 28} (1997), 1158–1172.


\bibitem{Marek}
I. Marek,
\newblock{ Frobenius theory for positive operators: : Comparison theorems and applications},
\newblock {\em SIAM J. Appl. Math.}, {\bf 19} (1970), 607-628.

\bibitem{MD}
J. A. J. Metz and O. Diekmann,
\newblock {\em The Dynamics of Physiologically Structured Populations},
\newblock Springer, Berlin, (1986).

\bibitem{NI}
H. Nishiura and H. Inaba,
\newblock Discussion: Emergence of the concept of the basic reproduction number from mathematical demography,
\newblock  {\em J. Theoret. Biol.}, {\bf 244} (2007), 357–364.


\bibitem{Sch}
H.~H. Sch\"{a}fer,
\newblock {\em Banach lattices and positive operators},
\newblock Springer Verlag, Berlin, (1974).

\bibitem{HT2}
H. R. Thieme,
\newblock Eigenvectors and eigenfunctionals of homogeneous order-preserving maps,
\newblock {\em http://arxiv.org/abs/1302.3905}.

\bibitem{HT}
H. R. Thieme,
\newblock Spectral bound and reproduction number for infinite-dimensional population structure and time heterogeneity,
\newblock {\em SIAM J. Appl. Math.}, {\bf 70} (2009), 188-211.

\bibitem{W}
G. F. Webb,
\newblock {\em Theory of Nonlinear Age-Dependent Population Dynamics},
\newblock Marcel Dekker, New York, (1985).


\end{thebibliography}
\end{document}